\documentclass[conference]{IEEEtran}
\IEEEoverridecommandlockouts
\usepackage{amsmath,amssymb,amsfonts}
\usepackage{algorithmic}
\usepackage{graphicx}
\usepackage{textcomp}
\usepackage{xcolor}
\usepackage{booktabs}
\usepackage{comment}
\usepackage{multirow}
\usepackage{tikz}

\usetikzlibrary{shapes.geometric, arrows.meta, positioning, fit, backgrounds}
\def\BibTeX{{\rm B\kern-.05em{\sc i\kern-.025em b}\kern-.08em
    T\kern-.1667em\lower.7ex\hbox{E}\kern-.125emX}}
\begin{document}

\title{Agentic Knowledge Distillation: Autonomous Training of Small Language Models for SMS Threat Detection}

\author{
\IEEEauthorblockN{Adel ElZemity}
\IEEEauthorblockA{\textit{School of Computing} \\
\textit{University of Kent}\\
Canterbury, United Kingdom \\
ae455@kent.ac.uk}
\and
\IEEEauthorblockN{Joshua Sylvester}
\IEEEauthorblockA{\textit{School of Computing} \\
\textit{University of Kent}\\
Canterbury, United Kingdom \\
jrs71@kent.ac.uk}
\and
\IEEEauthorblockN{Budi Arief}
\IEEEauthorblockA{\textit{School of Computing} \\
\textit{University of Kent}\\
Canterbury, United Kingdom \\
b.arief@kent.ac.uk}
\and
\IEEEauthorblockN{Rogério De Lemos}
\IEEEauthorblockA{\textit{School of Computing} \\
\textit{University of Kent}\\
Canterbury, United Kingdom \\
r.delemos@kent.ac.uk}
}

\maketitle

\begin{abstract}
SMS-based phishing (smishing) attacks have surged, yet training effective on-device detectors requires labelled threat data that quickly becomes outdated. To deal with this issue, we present Agentic Knowledge Distillation, which consists of a powerful LLM acts as an autonomous teacher that fine-tunes a smaller student SLM, deployable for security tasks without human intervention. The teacher LLM autonomously generates synthetic data and iteratively refines a smaller on-device student model until performance plateaus. We compare four LLMs in this teacher role (Claude Opus 4.5, GPT 5.2 Codex, Gemini 3 Pro, and DeepSeek V3.2) on SMS spam/smishing detection with two student SLMs (Qwen2.5-0.5B and SmolLM2-135M). Our results show that performance varies substantially depending on the teacher LLM, with the best configuration achieving 94.31\% accuracy and 96.25\% recall. We also compare against a Direct Preference Optimisation (DPO) baseline that uses the same synthetic knowledge and LoRA setup but without iterative feedback or targeted refinement; agentic knowledge distillation substantially outperforms it (e.g. 86--94\% vs 50--80\% accuracy), showing that closed-loop feedback and targeted refinement are critical. These findings demonstrate that agentic knowledge distillation can rapidly yield effective security classifiers for edge deployment, but outcomes depend strongly on which teacher LLM is used.
\end{abstract}


\section{Introduction}

SMS-based attacks have become a critical cyber security threat. Smishing (SMS phishing) attacks increased by over 300\% in recent years, with attackers exploiting mobile messaging to deliver phishing links, fake banking alerts, cryptocurrency scams, and social engineering attempts~\cite{smishing_threat}. Unlike email spam, SMS messages bypass many traditional security filters and benefit from higher user trust, and the recipients are more likely to click on the link contained in the SMS, making them particularly effective attack vectors. The challenge for defenders is that SMS spam evolves rapidly, requiring security models that can adapt quickly to new threat patterns.

Deploying large language models (LLMs) for on-device SMS filtering faces significant challenges: high computational costs, latency constraints, and privacy concerns when processing personal messages. A natural way to address these limitations is to use small language models (SLMs).

Following recent taxonomies in the field \cite{wang2024slmsurvey, zhang2024slmmeasure}, we define an SLM as a neural language model with a parameter count typically under 10 billion (e.g., 100M to 7B), characterised by its ability to perform inference on local, resource-constrained consumer hardware without reliance on cloud infrastructure. Conversely, we define LLMs as models exceeding this parameter threshold that typically exhibit emergent generalist capabilities but necessitate massive distributed memory for execution. Within this architectural boundary, we specifically adopt the functional perspective of Belcak et al. \cite{belcak2025smalllanguagemodelsfuture}, identifying an SLM as a model capable of serving a single user’s agentic workloads with practically acceptable latency on their local device.

SLMs offer a practical alternative for edge deployment, but they require task-specific fine-tuning, a process that traditionally demands substantial human expertise and labelled data, which may not capture emerging threat patterns.
Knowledge distillation~\cite{hinton2015distilling} addresses this gap.
A teacher LLM (typically an LLM with strong task capability) transfers its knowledge to a student SLM, a smaller model that is trained to replicate the teacher LLM's behaviour and can then be deployed where the teacher LLM cannot. In our setting, the student is an SLM for on-device SMS filtering, and the teacher is an LLM that provides the task knowledge. However, conventional distillation requires human engineers to design data pipelines, analyse errors, and iterate on training strategies, a slow process ill-suited to rapidly evolving threats.

As a non-agentic baseline, we also consider Direct Preference Optimisation (DPO), a recently proposed alignment and fine-tuning paradigm that optimises a model directly from preference pairs without an explicit reinforcement learning loop. In our context, DPO provides a useful point of comparison for assessing whether the additional complexity of agentic knowledge distillation is justified. 
Rather than using an autonomous LLM agent to iteratively generate data and refine the student, DPO fine-tunes the student SLM in a single-stage procedure using synthetic preference data derived from teacher LLM outputs. This allows us to isolate the benefits of iterative, agent-driven adaptation over a simpler, static fine-tuning pipeline.

We propose a novel \textit{Agentic Knowledge Distillation} approach, which is an automated framework where the LLM itself serves as an autonomous machine learning (ML) engineer. Given a task specification and evaluation criteria, the LLM agent generates synthetic training data from its knowledge (including patterns of phishing, smishing, and social engineering attacks), executes fine-tuning on the student SLM, evaluates performance against a separate, teacher-generated synthetic validation set, and iterates until performance plateaus. This ensures the optimisation process remains entirely self-contained and independent of human-labelled data, and enables rapid adaptation to new threat categories.

For our teacher LLM We choose four models ranked within the top 10 on the Artificial Analysis Intelligence Index (as of January 2026)~\cite{artificialanalysis2025}: 
\begin{itemize}
    \item  Claude Opus 4.5 from Anthropic~\cite{claudeopus45}, 
    \item  GPT 5.2 Codex from OpenAI~\cite{gpt52codex}, 
    \item  Gemini 3 Pro from Google~\cite{gemini3pro}, 
    \item  DeepSeek V3.2 \cite{deepseekv32}. 
\end{itemize}
We compare these four as teacher LLMs to test whether leading general-purpose LLMs transfer  knowledge effectively to SLMs. For student SLMs, we use Qwen2.5-0.5B-Instruct (494M parameters) and SmolLM2-135M-Instruct (135M parameters).
These can be deployed on consumer hardware with acceptable latency, span a useful range of capacity to test how student SLM size interacts with teacher LLM quality, and are instruction-tuned for consistent task formatting.

We pair each teacher LLM with each student SLM (eight configurations), using the SMS Spam Collection as a strictly held-out test set~\cite{almeida2011sms}. Each teacher LLM generates synthetic training data, fine-tunes the student SLM with Low-Rank Adaptation (LoRA), and iterates based only on aggregate evaluation metrics, never on raw test examples. Our results reveal that teacher LLM capability significantly impacts student SLM performance, with substantial variation in final accuracy and precision-recall balance across teacher LLMs.

Our contributions are: 
\begin{enumerate}
    \item We introduce an agentic knowledge distillation framework for autonomous fine-tuning of deployable student SLMs using a teacher LLM.
    \item We provide a study showing that teacher LLM choice is important for the effectiveness of agentic knowledge distillation.
\end{enumerate}

The rest of the paper is organised as follows. 
Section~\ref{sec:RelatedWork} reviews related work on SMS spam detection, knowledge distillation, synthetic data generation, and LLM agents. 
Section~\ref{sec:Methodology} describes our methodology, including the agentic workflow and strict train-test separation. 
Section~\ref{sec:Experiments} outlines the experimental setup.
The results are discussed in Section~\ref{sec:Results}.
Section~\ref{sec:Discussion} discusses implications and limitations.
Section~\ref{sec:Conclusions} concludes the paper and presents directions of future research. 
The system prompt given to each teacher LLM is in the Appendix.

\section{Related Work}
\label{sec:RelatedWork}

This section reviews related work on SMS spam detection, knowledge distillation, and LLM-based synthetic data generation, and situates our approach within recent agentic and preference-based training methods.

\textbf{SMS Spam and Smishing Detection.} SMS spam detection has been studied extensively as both a text classification problem and a security challenge~\cite{almeida2011sms}. Traditional approaches used Naive Bayes, SVM, and rule-based filters~\cite{sms_spam_survey}. However, attackers continuously adapt their tactics, using URL shorteners, homoglyphs, and social engineering to evade detection. Recent work has explored deep learning approaches, but these typically require large labelled datasets that become outdated as attack patterns evolve~\cite{Gadde2021}. Our approach addresses this by leveraging LLM knowledge of attack patterns to generate diverse synthetic training data without requiring manual labelling of emerging threats.

\textbf{Knowledge Distillation.} Knowledge distillation, introduced by Hinton et al.~\cite{hinton2015distilling}, transfers knowledge from large models to smaller ones through soft label matching. Recent work has extended this to LLMs, with approaches like DistilBERT~\cite{sanh2019distilbert} and TinyBERT~\cite{jiao2020tinybert} demonstrating effective compression. Our work differs by eliminating the need for a shared training dataset: the teacher LLM generates data directly from its parametric knowledge.

\textbf{Synthetic Data Generation.} LLMs have shown remarkable capability in generating training data for various tasks. Self-Instruct~\cite{wang2023selfinstruct} demonstrated that models can generate instruction-following data to improve their own performance, while Stanford Alpaca~\cite{alpaca} used this approach to create instruction-tuned models. We extend this concept to cross-model distillation for security applications, where a large model generates synthetic attack and benign examples to train a smaller, deployable detector.

\textbf{LLM Agents.} Recent advances in LLM agents~\cite{autogpt,langchain} have enabled autonomous task completion through tool use and iterative reasoning. Our work applies this agentic paradigm to the ML engineering process itself, treating model training as a task the LLM can autonomously complete. Combined with parameter-efficient fine-tuning methods like LoRA~\cite{hu2022lora}, this enables rapid iteration on consumer hardware.

\textbf{Preference Optimisation.} Preference-based fine-tuning has emerged as an alternative to reinforcement learning from human feedback (RLHF) for aligning language models with desired behaviours~\cite{dpo}. Direct Preference Optimisation (DPO)~\cite{dpo} formulates preference learning as a supervised objective over paired outputs, avoiding the need for an explicit reward model or online policy optimisation. DPO and related methods have been successfully applied to instruction following and helpfulness alignment in general-purpose LLMs~\cite{dpo}. We apply DPO in a security classification context by preference-tuning the student SLM on teacher LLM generated preferences over candidate classifications. This serves as a strong baseline that relies on static synthetic supervision without any failure pattern-driven data generation.

\section{Methodology}
\label{sec:Methodology}

Given a teacher LLM $T$, a student SLM $S_\theta$, and a task description $D$, our goal is to fine-tune the student SLM for detecting SMS spam and smishing using only synthetic data generated by the teacher LLM $T$. 
The teacher LLM must autonomously generate all training data and orchestrate the fine-tuning process without access to any test data. 

\begin{figure}[htbp]
    \centering
    \includegraphics[width=0.95\linewidth]{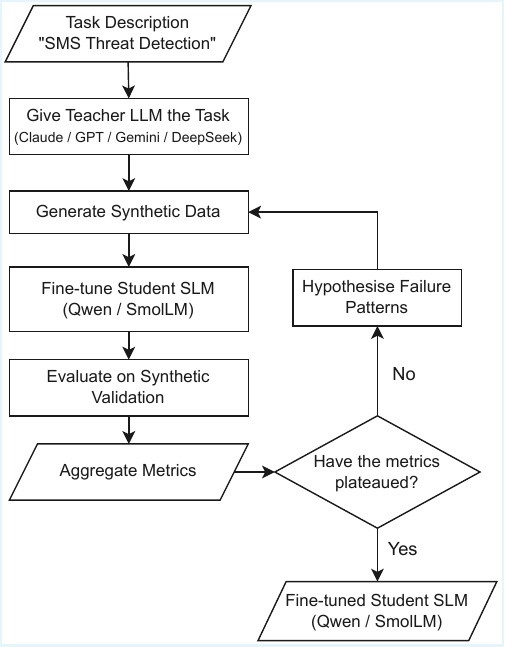}
    \caption{Agentic knowledge distillation workflow. The teacher LLM generates both synthetic training and synthetic validation sets. The conditional loop uses metrics derived strictly from the synthetic validation set. The final evaluation on the human-labelled test set on the outputted fine-tuned student SLM.}
    \label{fig:workflow}
\end{figure}

\subsection{Workflow Overview}

Figure~\ref{fig:workflow} presents the agentic knowledge distillation workflow used to fine-tune a student SLM for SMS threat detection under the supervision of a teacher LLM. 
The workflow is formulated as an autonomous, closed-loop optimisation process, where the teacher LLM iteratively generates data, fine-tunes the student SLM, evaluates performance, and refines its strategy until performance converges.

The process begins with a task specification (``SMS Threat Detection'') that is provided to the teacher LLM (e.g., Claude, GPT, Gemini, or DeepSeek), as shown at the top of Figure~\ref{fig:workflow}. 
Each teacher LLM is given an identical system prompt defining the mission objective and the structure of the iterative loop (see Appendix). 
This prompt specifies:
\begin{enumerate}
    \item the objective of fine-tuning the student SLM for SMS threat detection, and
    \item an iterative cycle comprising synthetic data generation, fine-tuning, evaluation, error analysis, and refinement.
\end{enumerate}

Given the task, the teacher LLM first generates a synthetic training dataset $\mathcal{F}$, which is used to fine-tune the student SLM (e.g., Qwen2.5-0.5B or SmolLM2-135M). 
The fine-tuned student is then evaluated on a held-out synthetic validation set $\mathcal{V}$, which is generated once at the start of the workflow and remains fixed across iterations. 
The resulting performance metrics are aggregated and fed back to the teacher LLM.

The teacher LLM then determines whether performance has plateaued. 
If the validation metrics are deemed to have converged, the workflow terminates and the final fine-tuned student SLM is produced as the output of the pipeline. 
Otherwise, the teacher LLM performs error analysis by hypothesising failure patterns (e.g., systematic misclassification of particular phishing styles or obfuscation strategies) and generates additional targeted synthetic data to address these weaknesses. 
This newly generated data is incorporated into the training set, and the fine-tuning and evaluation loop repeats until the stopping criterion is met.

Throughout this process, the teacher LLM operates as an autonomous agent with full terminal access, enabling it to install libraries, generate datasets, execute fine-tuning scripts, and invoke the external evaluation system. 
Crucially, the optimisation loop is driven solely by synthetic validation metrics. 
The real-world SMS Spam Collection dataset is strictly \emph{air-gapped} from the agentic workflow and is used only for final, post-hoc evaluation to assess real-world generalisation.

\subsubsection{Synthetic Data Generation}

The teacher LLM generates a set of samples from its own knowledge of the domain then writes a script that would automate generating a larger synthetic training dataset
\[
\mathcal{F} = \{(\tilde{x}_i, \tilde{y}_i)\}_{i=1}^{N},
\]
where $\tilde{x}_i$ is a synthetic SMS message and $\tilde{y}_i \in \{0,1\}$ denotes a benign or malicious label assigned by construction.

Synthetic messages are generated from the teacher LLM’s knowledge of the domain and cover a diverse range of attack categories, including phishing, smishing, fake delivery alerts, cryptocurrency scams, lottery fraud, and premium-rate abuse.
Benign messages include casual conversations, workplace communications, and legitimate service notifications.

To avoid class imbalance, the agent is instructed to generate approximately equal numbers of spam and ham messages ($50/50$ split). 

\subsubsection{Student SLM Fine-Tuning with LoRA}

The student SLM is fine-tuned using Low-Rank Adaptation (LoRA). 
Rather than updating all the model parameters, LoRA injects small trainable matrices into selected layers while keeping the base model frozen. 
This enables efficient training on consumer hardware.

Let $S_\theta$ denote the student SLM where $\theta$ are the trainable LoRA parameters. 
Training minimises binary cross-entropy loss over the synthetic dataset:
\[
\mathcal{L}(\theta) =
- \frac{1}{N} \sum_{i=1}^{N}
\left[
\tilde{y}_i \log S_\theta(\tilde{x}_i)
+ (1 - \tilde{y}_i)\log(1 - S_\theta(\tilde{x}_i))
\right].
\]


\subsubsection{Internal Synthetic Validation vs.\ External Testing}

\textbf{1) Internal Loop:} For the iterative refinement, the Teacher LLM evaluates the Student SLM on the synthetic validation set $\mathcal{V}$. This provides the feedback vector $M_t$ used to hypothesise failure patterns. Since $\mathcal{V}$ is generated by the Teacher LLM, the ``ground truth'' labels are known without human intervention. At iteration $t$, the teacher LLM receives a metric vector
\[
M_t = (\text{Acc}_t, \text{Prec}_t, \text{Rec}_t, \text{FP}_t, \text{FN}_t).
\]

\textbf{2) External Evaluation:} The human-labelled SMS Spam Collection is used solely as a final ``sanity check'' after the agentic process has terminated. These external metrics are reported in our Results tables~\ref{tab:baseline},~\ref{tab:results_dpo},~\ref{tab:results} but are never fed back to the Teacher LLM.

\subsubsection{Error Analysis and Targeted Refinement}

Based solely on aggregate metrics, the teacher LLM hypothesises likely failure patterns as shown in Figure~\ref{fig:workflow}. 
For example, a high false-positive rate suggests that legitimate service messages may be misclassified as spam. 
In contrast, a high false-negative rate indicates insufficient coverage of subtle or short-form scams.

The teacher LLM then generates additional synthetic examples specifically targeting these hypothesised weaknesses. 
This process resembles hard negative mining, but differs in that no actual misclassified examples are observed~\cite{robinson2021contrastive}. 
Instead, the teacher LLM relies on domain knowledge to construct targeted refinements. 

The refined dataset is used for further LoRA fine-tuning, and the evaluation-feedback loop repeats until performance plateaus.

\subsection{Evaluation of Framework}

To evaluate whether the performance gains observed in our experiments arise from the proposed agentic knowledge distillation framework itself, rather than merely from access to high-quality synthetic data, we introduce a teacher LLM-generated Direct Preference Optimisation (DPO) baseline. This baseline is designed to isolate the contribution of closed-loop, agent-driven refinement by providing a strong non-agentic alternative that uses the same teacher models, the same synthetic knowledge source, and the same fine-tuning setup, but without iterative feedback or targeted data generation.

Specifically, we fine-tune the student SLM using Direct Preference Optimisation (DPO)~\cite{dpo} on preference data generated by a teacher LLM and compare the resulting performance against the full agentic knowledge distillation framework. This enables a controlled comparison between static preference-based fine-tuning and the proposed autonomous, closed-loop training process.

For each teacher LLM $T$, we construct a synthetic preference dataset. Each sample comprises a prompt $x$ (an SMS message) and a pair of candidate responses $(y^{+}, y^{-})$, where $y^{+}$ represents the preferred classification output and $y^{-}$ a less suitable alternative, as determined by the teacher LLM. Preferences are generated using the same domain knowledge, attack categories, and class balance assumptions as in the agentic knowledge distillation setting, ensuring that both approaches are trained under comparable supervision signals.

The student SLM $S_\theta$ is then fine-tuned using DPO under an identical LoRA configuration to the agentic method. All architectural choices, LoRA ranks, scaling factors, learning rates, batch sizes, and training budgets are held constant to ensure a fair comparison between training regimes. Crucially, the DPO baseline is trained in a single offline stage and does not receive any iterative metric feedback, perform failure analysis, or generate targeted refinements.

Formally, DPO optimises the student SLM to increase the likelihood of preferred outputs relative to non-preferred ones under the teacher-induced preference distribution. In contrast to the agentic workflow, this approach does not involve hypothesis generation, error analysis, or adaptive hard-negative synthesis. As a result, it provides a controlled non-agentic baseline that isolates the effect of preference-based fine-tuning from the benefits of autonomous closed-loop reasoning.

The DPO-trained models are evaluated using the same strictly held-out test set. This enables direct comparison between (i) zero-shot baselines, (ii) teacher LLM-generated DPO fine-tuning, and (iii) the proposed agentic knowledge distillation framework.

\subsection{Summary}

The resulting system forms a closed-loop agentic knowledge distillation workflow in which the teacher LLM autonomously generates data, fine-tunes the student SLM, and iteratively improves performance using only metric-level feedback. 
Figure~\ref{fig:workflow} summarises this workflow. 

The strict separation between data generation and evaluation ensures that observed improvements reflect genuine generalisation rather than overfitting.
We additionally compare against a teacher LLM-generated DPO baseline that leverages the same synthetic knowledge and identical LoRA configurations, but lacks iterative feedback and targeted refinement. 
This comparison highlights the contribution of agentic reasoning and closed-loop adaptation beyond static preference-based fine-tuning.

\section{Experiments}
\label{sec:Experiments}

We design our experiments to evaluate (i) the effectiveness of agentic knowledge distillation for SMS threat detection, (ii) the impact of teacher LLM choice, and (iii) the benefit of closed-loop refinement compared to static synthetic fine-tuning.

\subsection{Models}

We evaluate four teacher LLMs: Claude Opus 4.5 \cite{claudeopus45}, GPT 5.2 Codex \cite{gpt52codex}, Gemini 3 Pro \cite{gemini3pro}, and DeepSeek V3.2 \cite{deepseekv32}, all accessed via OpenRouter \cite{openrouter}. Each teacher is used to fine-tune two student SLMs: Qwen2.5-0.5B-Instruct~\cite{yang2024qwen25} and SmolLM2-135M-Instruct~\cite{allal2025smollm2}, using identical LoRA configurations across all runs.

\subsection{Datasets and Evaluation Protocol}

For evaluation, we use the SMS Spam Collection \cite{almeida2011sms}, a public corpus of 5,574 SMS messages aggregating data from the Grumbletext UK forum (425 spam), NUS SMS Corpus (3,375 ham), Caroline Tag's PhD thesis (450 ham), and SMS Spam Corpus v.0.1. 
To ensure a fair and interpretable evaluation, we construct a balanced test subset of 1,494 messages (747 spam, 747 ham), and remove the remaining ham messages from the evaluation set. 
This prevents the reported metrics from being dominated by the majority ``ham'' class, which would otherwise inflate accuracy and obscure failure modes on the minority spam class. 
Balancing the evaluation set enables more meaningful comparison across models and training regimes, particularly for recall and F1 score, which are critical for assessing threat detection performance. 
Crucially, the agentic training workflow never had access to the content of the test data, which remained strictly held out and air-gapped from the synthetic data generation and fine-tuning process.

\subsection{Experimental Methods}

We conduct the following experiments:

\begin{itemize}
    \item \textbf{Zero-shot baselines.} Both student SLMs are evaluated without any fine-tuning using zero-shot prompting to establish a lower bound on performance.
    
    \item \textbf{DPO with synthetic data.} For each teacher LLM, we generate a synthetic preference dataset of 10,000 samples and fine-tune each student SLM using Direct Preference Optimisation (DPO) with LoRA. This setting isolates the effect of synthetic data generation without closed-loop refinement.
    
    \item \textbf{Agentic knowledge distillation.} For each teacher LLM and student SLM pair, we run the full closed-loop agentic workflow, in which the teacher autonomously generates synthetic training and validation data, analyses errors, and iteratively refines the training set until validation performance plateaus.
\end{itemize}

This yields a total of $4 \times 2 \times 3$ experimental configurations (four teachers, two students, and three experimental methods).

\subsection{Implementation Details}

All experiments are conducted on a Mac Mini M4 with 16GB unified memory. For fairness, we use identical optimisation settings, LoRA ranks, learning rates, and stopping criteria across all experimental conditions. We record token usage for each agentic run to characterise computational cost.
For the LoRA configuration we use a rank of 32, scaling factor of 64, learning rate of $5 \times 10^{-5}$, and batch size of 8. 

\section{Results}
\label{sec:Results}

This section presents results in three stages. 
We first report zero-shot performance of the student SLMs without any fine-tuning to establish the initial baseline. 
We then evaluate a teacher LLM-generated Direct Preference Optimisation (DPO) baseline, which uses a single offline round of synthetic preference data without iterative refinement. 
Finally, we present results for the proposed agentic knowledge distillation framework, analysing performance across different teacher LLMs and student SLMs, and comparing against both the zero-shot and DPO baselines.

\begin{table}[htbp]
\caption{Baseline Performance (No Fine-tuning)}
\begin{center}
\begin{tabular}{lcccc}
\toprule
\textbf{Model} & \textbf{Acc.} & \textbf{Prec.} & \textbf{Recall} & \textbf{F1} \\
\midrule
Qwen2.5-0.5B & 49.80\% & 28.57\% & 0.27\% & 0.53\% \\
SmolLM2-135M & 46.85\% & 46.74\% & 45.11\% & 46.00\% \\
\bottomrule
\end{tabular}
\label{tab:baseline}
\end{center}
\end{table}

Both student SLMs were first evaluated without fine-tuning using zero-shot prompting (Table~\ref{tab:baseline}). 
Qwen2.5-0.5B exhibits an extreme bias towards predicting ``ham'', correctly identifying only 2 out of 747 spam messages, resulting in near-zero recall and F1 score. 
SmolLM2-135M performs close to random chance, indicating limited inherent discriminative capability for the task in the absence of task-specific fine-tuning.

\begin{table*}[htbp]
\caption{Final Performance by Data-Generating Teacher LLM (DPO)}
\begin{center}
\begin{tabular}{llccccc}
\toprule
\textbf{Teacher LLM} & \textbf{Student SLM} & \textbf{Acc.} & \textbf{Prec.} & \textbf{Recall} & \textbf{F1} & \textbf{Tokens} \\
\midrule
\multirow{2}{*}{Claude Opus 4.5} 
 & Qwen2.5-0.5B 
 & 52.74\% 
 & 72.52\% 
 & 52.74\% 
 & 39.45\% 
 & \multirow{2}{*}{2142.48K} \\
 & SmolLM2-135M 
 & \textbf{80.25\%} 
 & \textbf{81.81\%}
 & \textbf{80.25\%}
 & \textbf{80.01\%} 
 &  \\
\midrule
\multirow{2}{*}{GPT 5.2 Codex} 
 & Qwen2.5-0.5B 
 & 51.34\% 
 & 57.30\% 
 & 51.34\% 
 & 38.86\% 
 & \multirow{2}{*}{1721.52K} \\
 & SmolLM2-135M 
 & 52.61\% 
 & 70.57\% 
 & 52.61\% 
 & 39.38\% 
 &  \\
\midrule
\multirow{2}{*}{Gemini 3 Pro} 
 & Qwen2.5-0.5B 
 & 52.54\% 
 & 68.93\% 
 & 52.54\% 
 & 39.44\% 
 & \multirow{2}{*}{1839.37K} \\
 & SmolLM2-135M 
 & 68.27\% 
 & 77.33\% 
 & 68.27\% 
 & 65.41\% 
 &  \\
\midrule
\multirow{2}{*}{DeepSeek V3.2}
 & Qwen2.5-0.5B
 & 52.48\%
 & 68.09\%
 & 52.48\%
 & 39.40\%
 & \multirow{2}{*}{1892.48K} \\
 & SmolLM2-135M
 & 76.44\%
 & 78.44\%
 & 76.44\%
 & 76.02\%
 &  \\
\bottomrule
\end{tabular}
\label{tab:results_dpo}
\end{center}
\end{table*}

Table~\ref{tab:results_dpo} reports results for the teacher LLM-generated DPO baseline. 
While DPO fine-tuning yields modest improvements over the zero-shot baselines in some settings, performance remains limited and highly variable across teacher--student pairs. 
In particular, Qwen2.5-0.5B remains close to chance performance under DPO across all teachers, whereas SmolLM2-135M benefits more substantially, achieving up to 80.25\% accuracy when trained on Claude-generated preference data. 
These results highlight the limitations of single-stage, static preference fine-tuning in the absence of iterative error-driven refinement.

\begin{table*}[htbp]
\caption{Final Performance by Agentic Knowledge Distillation Teacher LLM}
\begin{center}
\begin{tabular}{llcccccc}
\toprule
\textbf{Teacher LLM} & \textbf{Student SLM} & \textbf{Acc.} & \textbf{Prec.} & \textbf{Recall} & \textbf{F1} & \textbf{Tokens} & \textbf{Time} \\
\midrule
\multirow{2}{*}{Claude Opus 4.5} & Qwen2.5-0.5B & \textbf{94.31\%} & 92.65\% & \textbf{96.25\%} & \textbf{94.42\%} & 27.91K & $\sim$7 min \\
 & SmolLM2-135M & \textbf{86.28\%} & 80.38\% & 95.98\% & \textbf{87.00\%} & 30.56K & $\sim$6 min \\
\midrule
\multirow{2}{*}{GPT 5.2 Codex} & Qwen2.5-0.5B & 71.08\% & \textbf{98.76\%} & 42.70\% & 59.64\% & 26.37K & $\sim$6 min \\
 & SmolLM2-135M & 59.71\% & 55.38\% & \textbf{99.87\%} & 71.26\% & 24.37K & $\sim$5 min \\
\midrule
\multirow{2}{*}{Gemini 3 Pro} & Qwen2.5-0.5B & 85.21\% & 79.29\% & 95.31\% & 86.58\% & 14.83K & $\sim$9 min \\
 & SmolLM2-135M & 80.46\% & 72.73\% & 97.46\% & 83.31\% & 16.78K & $\sim$8 min \\
\midrule
\multirow{2}{*}{DeepSeek V3.2} & Qwen2.5-0.5B & 92.10\% & 91.77\% & 92.50\% & 92.13\% & 31.2K & $\sim$8 min \\
 & SmolLM2-135M & 86.21\% & \textbf{88.70}\% & 83.00\% & 85.75\% & 32.1K & $\sim$7 min \\
\bottomrule
\end{tabular}
\label{tab:results}
\end{center}
\end{table*}

\begin{table}[htbp]
\caption{Accuracy Improvement by Agentic Knowledge Distillation Teacher LLM ($\Delta$ from Baseline)}
\begin{center}
\begin{tabular}{lcc}
\toprule
\textbf{Teacher LLM} & \textbf{Qwen2.5-0.5B} & \textbf{SmolLM2-135M} \\
\midrule
Claude Opus 4.5 & \textbf{+44.51 pp} & \textbf{+39.43 pp} \\
DeepSeek V3.2 & +42.30 pp & +39.36 pp \\
Gemini 3 Pro & +35.41 pp & +33.61 pp \\
GPT 5.2 Codex & +21.28 pp & +12.86 pp \\
\bottomrule
\end{tabular}
\label{tab:improvement}
\end{center}
\end{table}

Table~\ref{tab:results} presents the final performance across all teacher LLM and student SLM combinations, and Table~\ref{tab:improvement} shows the accuracy improvements from baseline in Table~\ref{tab:baseline}. 
The results reveal substantial variation in the effectiveness of agentic knowledge distillation across teacher LLMs.

Comparing Table~\ref{tab:results_dpo} and Table~\ref{tab:results} reveals a clear and consistent performance gap between the teacher LLM-generated DPO baseline and the proposed agentic knowledge distillation approach. While DPO preference-tuning yields modest improvements over zero-shot baselines, performance remains limited, with accuracies typically ranging from 50--80\% and substantially lower F1 scores, particularly for the Qwen2.5-0.5B student SLM. In contrast, agentic knowledge distillation achieves markedly higher and more stable performance across all teacher LLM and student SLM pairs, reaching up to 94.31\% accuracy and 94.42\% F1 (Table~\ref{tab:results}). 

\textbf{Claude Opus 4.5} achieved the best overall results, producing well-balanced classifiers with high accuracy (86--94\%), strong precision (80--93\%), and excellent recall (96\%). The hard negative mining phase proved particularly effective, i.e., for Qwen2.5-0.5B, false positives decreased from 207 to 57 after refinement.

\textbf{GPT 5.2 Codex} produced models with severe precision-recall imbalances. For Qwen2.5-0.5B, the fine-tuned model achieved very high precision (98.76\%) but poor recall (42.70\%), missing more than half of spam messages. For SmolLM2-135M, the opposite occurred: near-perfect recall (99.87\%) but low precision (55.38\%), flagging nearly half of legitimate messages as spam. Both configurations resulted in lower overall accuracy than Claude-trained models.

\textbf{DeepSeek V3.2} achieved strong, well-balanced results, with 86--92\% accuracy, precision and recall both in the 83--93\% range. For Qwen2.5-0.5B, it reached 92.10\% accuracy and 92.13\% F1, and for SmolLM2-135M, 86.21\% accuracy and 85.75\% F1. Its balanced precision-recall contrasts with GPT's severe imbalances and places it among the most effective teacher LLMs.

\textbf{Gemini 3 Pro} achieved intermediate results, ranking behind Claude Opus 4.5 and DeepSeek V3.2. It produced reasonably balanced classifiers with 80--85\% accuracy, moderate precision (73--79\%), and high recall (95--97\%). Notably, Gemini 3 pro was the most token-efficient teacher LLM, using only 14--17K tokens compared to 24--33K for the other models. However, its lower precision compared to Claude Opus 4.5 and DeepSeek V3.2 suggests less effective hard negative mining for reducing false positives.

The GPT 5.2 Codex results highlight a critical failure pattern in agentic knowledge distillation: when the teacher LLM generates imbalanced or insufficiently diverse synthetic data, the student SLM develops biased decision boundaries. The Qwen2.5-0.5B model trained by GPT 5.2 Codex became overly conservative (high precision, low recall), while the SmolLM2-135M model became overly aggressive (high recall, low precision).

\section{Discussion}
\label{sec:Discussion}

The comparison between the DPO baseline and agentic knowledge distillation, as observed in Section~\ref{sec:Results}, provides insight into the source of the performance gains. Although DPO leverages the same teacher LLMs, synthetic domain knowledge, and identical LoRA configurations, its substantially lower accuracy and F1 scores (Table~\ref{tab:results_dpo}) indicate that static preference-based optimisation alone is insufficient to induce robust decision boundaries. In comparison, the agentic method’s iterative use of aggregate evaluation feedback to hypothesise failure patterns and generate targeted synthetic refinements consistently produces well-balanced classifiers (Table~\ref{tab:results}). This suggests that the primary benefit does not arise from the data source itself, but from the closed-loop reasoning and adaptive training dynamics enabled by agentic knowledge distillation, which are critical for closing precision--recall gaps and achieving strong generalisation.

Our results demonstrate that teacher LLM selection is a critical factor in agentic knowledge distillation. The performance gap between teacher LLMs is substantial: Claude Opus 4.5 achieved 86--94\% accuracy, DeepSeek V3.2 achieved 86--92\% with balanced precision-recall, Gemini 3 Pro achieved 80--85\%, and GPT 5.2 Codex achieved only 60--71\%. This 25+ percentage point spread suggests that not all LLMs are equally effective as autonomous machine learning (ML) engineers.

Several factors may explain these differences. 
First, the coverage and structure of the synthetic training data appear to vary across teacher LLMs. 
The stronger performance of Claude Opus 4.5 and DeepSeek V3.2 suggests that their generated datasets more consistently captured a broad range of spam strategies (e.g., phishing, scams, obfuscation) as well as legitimate conversational patterns, leading to more robust decision boundaries. 
In contrast, Gemini 3 Pro has high recall but lower precision which indicates that its synthetic data may have insufficiently represented challenging ham examples, causing the student models to over-generalise spam patterns and produce more false positives. 
GPT 5.2 Codex exhibits extreme precision–recall imbalances, pointing to systematic biases in the synthetic data generation process that skewed the student models toward overly conservative or overly aggressive classification regimes. 
Second, the effectiveness of the teacher LLM’s error analysis and hard negative mining differs: Claude Opus 4.5 and DeepSeek V3.2 appear more capable of identifying recurring failure modes and generating targeted counterexamples, leading to more stable and well-calibrated classifiers.

The precision-recall imbalances observed with GPT 5.2 Codex are particularly instructive. A spam detector  (Qwen2.5-0.5B trained by GPT 5.2 Codex) with 98.76\% precision but 42.70\% recall would miss most spam in practice, despite appearing ``precise.'' Conversely, a detector (SmolLM2-135M trained by GPT 5.2 Codex) with 99.87\% recall but 55.38\% precision would flag too many legitimate messages. These failure patterns highlight that agentic knowledge distillation can fail dramatically if the teacher LLM cannot generate balanced, high-quality synthetic data.

Interestingly, Gemini 3 Pro was the most token-efficient teacher LLM (14--17K tokens vs. 25--33K for others), yet was outperformed by Claude Opus 4.5 and DeepSeek V3.2. This suggests that token count alone does not determine success: the quality of reasoning and data generation matters more than quantity.

\textbf{Security Implications.} From a cyber security perspective, agentic knowledge distillation offers both opportunities and risks. On the defensive side, this approach enables rapid creation of threat detectors without requiring labelled datasets of emerging attack patterns: the LLM's knowledge of phishing tactics, social engineering, and fraud schemes can be distilled into lightweight, on-device models. This is particularly valuable for SMS security, where privacy concerns make cloud-based filtering undesirable. The best-performing configuration (Claude Opus 4.5 + Qwen2.5-0.5B) achieved 96.25\% recall, meaning it catches the vast majority of malicious messages while maintaining 92.65\% precision to avoid blocking legitimate communications.

However, the same technique could potentially be misused to train models for generating convincing phishing messages or evading detection. The observed variation in teacher LLM effectiveness may also apply to adversarial applications, though we note that generating effective attacks is likely harder than detecting them.

A key advantage of this framework is the resolution of the label scarcity paradox. Because the teacher LLM generates its own validation ``ground truth,'' the iterative feedback loop does not require a human-labelled oracle. The system is autonomous, requiring only a high-level task description to begin self-improvement.

\textbf{Limitations.} Performance is bounded by the teacher LLM's knowledge of the task domain: emerging attack patterns not represented in the LLM's training data may be missed. Additionally, while the agent never accesses test data content, it requires aggregate evaluation metrics on the synthetic validation set to guide iteration. The quality of synthetic data depends on the teacher LLM's knowledge, which may not cover all edge cases in real-world threats.

\section{Conclusions}
\label{sec:Conclusions}

We introduced agentic knowledge distillation, demonstrating that a large language model (LLM) can autonomously fine-tune a small language model (SLM) for security-critical tasks like SMS spam and smishing detection. Our evaluation of four teacher LLMs (Claude Opus 4.5, DeepSeek V3.2, Gemini 3 Pro, and GPT 5.2 Codex) reveals that teacher LLM selection critically impacts outcomes. Claude achieved the best results (86--94\% accuracy with balanced precision-recall), DeepSeek achieved strong results (86--92\% with balanced precision-recall), Gemini achieved moderate success (80--85\%), while GPT produced models with severe imbalances (60--71\%). The best configuration achieved 96.25\% recall for spam detection, demonstrating the potential for effective on-device mobile security. 

To contextualise this, we also evaluated a teacher LLM-generated Direct Preference Optimisation (DPO) baseline. While DPO fine-tuning improved performance over zero-shot baselines, it remained substantially lower than agentic knowledge distillation across all teacher LLM and student SLM pairs (50--80\% accuracy, lower F1 scores), highlighting that iterative feedback, hypothesis generation, and targeted data refinement are critical components of the agentic approach. 

Using SLMs as students is motivated by the need for local deployment: filtering runs on the device so that message content is never sent to external APIs (preserving privacy), latency remains acceptable for real-time use, and inference is feasible on consumer hardware where LLMs cannot run. These findings establish that careful teacher LLM selection is essential for successful agentic knowledge distillation, and that LLMs can serve as autonomous ML engineers for rapid development of security classifiers without requiring labelled threat data. The entire process requires only 15--33K tokens and completes in under 10 minutes on consumer hardware. 

This work suggests that agentic knowledge distillation is promising for security applications but requires careful teacher LLM selection. Future work should investigate adaptation to new threat categories, ensemble approaches combining multiple teacher LLMs, and longitudinal evaluation against evolving attack patterns.

\bibliographystyle{IEEEtran} 
\bibliography{main}

\appendix
\section{System Prompt}
\label{appendix:prompt}

The following system prompt was provided to all teacher LLM's:

\begin{quote}
\small
\textbf{Mission Objective:} Fine-tune [Student SLM] into a world-class SMS spam detector on this Mac Mini M4.

\textbf{Constraint:} Use the mlx-lm library to leverage the M4's GPU.

\textbf{Iterative Workflow:}
(1) \textbf{Baseline Evaluation}: Run the base model and calculate Accuracy, Recall, and Precision.
(2) \textbf{Agentic Knowledge Distillation}: Using your own knowledge, generate two distinct datasets: (a) a synthetic training set of 2,000+ examples, and (b) a held-out synthetic validation set ($\mathcal{V}$) of 500 examples for internal metrics. Both should be balanced 50/50 Spam/Ham and cover diverse categories: phishing, smishing, fake delivery alerts, crypto scams, and aggressive marketing.
(3) \textbf{Fine-Tuning}: Execute a LoRA fine-tune on the student SLM using the synthetic data.
(4) \textbf{Evaluation \& Feedback Loop}: Evaluate performance. Analyse aggregate error metrics and hypothesise which patterns may be causing errors. Generate targeted hard negatives based on these hypotheses. Repeat if performance hasn't plateaued.
(5) \textbf{Final Output}: Provide the final adapter weights and performance report.

\textbf{Resources Available:} Full terminal access, Python 3.14, and the MLX library. Begin.
\end{quote}

\end{document}